\documentclass[submission,copyright,creativecommons]{eptcs}
\providecommand{\event}{ACL2 Workshop 2023}
\usepackage{underscore, latexsym,pifont,amsmath,amstext,amssymb,amsfonts,makeidx,url}
\title{A Formalization of Finite Group Theory: Part III}
\author{David M.~Russinoff
\email{david@russinoff.com}
}
\def\titlerunning{A Formalization of Finite Group Theory: Part III}
\def\authorrunning{D.M. Russinoff}

\begin{document}
\maketitle

\begin{abstract}
This is the third and final installment of an exposition of an ACL2 formalization of finite group theory.  Part I covers groups and subgroups, cosets, normal subgroups, and quotient groups.  Part II extends the theory in the developmnent of group homomorphisms and direct products, which are applied in a proof of the Fundamental Theorem of Finite Abelian Groups.  The central topics of the present paper are the symmetric groups and the Sylow theorems, which pertain to subgroups of prime power order.  Since these theorems are based on the conjugation of subgroups, an example of a group action on a set, their presentation is preceded by a comprehensive treatment of group actions.  Our final result is mainly an application of the Sylow theorems: after showing that the alternating group of order 60 is simple (i.e., has no proper normal subgroup), we prove that no group of non-prime order less than 60 is simple.  The combined content of the {\tt groups} directory is a close approximation to that of an advanced undergraduate course taught by the author in 1976.
\end{abstract}

\section{Introduction}

This is the third and final installment of an exposition of an ACL2 formalization of finite group theory.  Part I \cite{part1}, which was presented at ACL2 2022, covers groups and subgroups, cosets, normal subgroups, and quotient groups.  Part II \cite{part2}, a companion paper in this workshop, extends the theory in the developmnent of group homomorphisms and direct products, which are applied in a proof of the Fundamental Theorem of Finite Abelian Groups.  The present paper is an account of four books of the directory {\tt projects/groups}---{\tt symmetric}, {\tt actions}, {\tt sylow}, and {\tt simple}---that have been appended to the seven books described in Parts I and II.  Part I is a prerequisite for a reading of this paper.  Parts II and III are largely independent, aside from several explicit references to Part II contained herein.

The central topics covered here are the symmetric groups {\tt (sym n)} (Section~\ref{sym}), consisting of the permutations of the list {\tt (0 1 2 ... n-1)}, and the Sylow theorems (Section~\ref{sylow}), which pertain to subgroups of prime power order.  Since these theorems are based on the conjugation of subgroups, an example of a group action on a set, their presentation is preceded by a comprehensive treatment of group actions (Section~\ref{actions}).  Our final result (Section~\ref{simple}) is mainly an application of the Sylow theorems: after showing that the alternating group {\tt (alt 5)}, of order 60, is simple (i.e., has no proper normal subgroup), we prove that no group of non-prime order less than 60 is simple.

\section{Symmetric Groups}\label{sym}

A {\it symmetric group} is the group of permutations of a given set under the operation of functional composition.  The study of these groups has important applications in diverse areas of mathematics and physics, such as combinatorics, Galois theory, and quantum mechanics.  They also provide a wide range of examples in group theory.  Since the elements of the underlying set are irrelevant to the group structure, it is common to focus on the permutations of an initial segment of the positive integers, $\{1, 2, \ldots, n\}$.  In our ACL2 formalization, it is more natural to consider the list {\tt (ninit n)} = {\tt (0 1 ... n-1)} of the first {\tt n} natural numbers.

\subsection{Definition of {\tt (sym n)}}

In \cite[Sec. 6]{part2}, we define a permutation of an arbitrary list:
\begin{small}
\begin{verbatim}
  (defun permutationp (l m)
    (if (consp l)
        (and (member-equal (car l) m)
             (permutationp (cdr l) (remove1-equal (car l) m)))
      (endp m)))
\end{verbatim}
\end{small}
In the special case of a list of distinct members, we have an equivalent formulation:
\begin{small}
\begin{verbatim}
  (defund permp (l m)
    (and (dlistp l) (dlistp m) (sublistp l m) (sublistp m l)))
  (defthmd permp-permutationp
    (implies (and (dlistp l) (dlistp m))
             (iff (permutationp l m) (permp l m))))
\end{verbatim}
\end{small}
The function {\tt perms} recursively constructs a list of all permutations of a dlist.  For a positive integer {\tt n}, the element list of the symmetric group {\tt (sym n)} is {\tt (slist n)}, the list of permutations of {\tt (ninit n)}:
\begin{small}
\begin{verbatim}
  (defund slist (n) (perms (ninit n)))
\end{verbatim}
\end{small}
A permutation {\tt x} in {\tt (slist n)} may be viewed as a bijection of {\tt (ninit n)}, which maps an integer {\tt k} to {\tt (nth k x)}.  The group operation is functional composition:
\begin{small}
\begin{verbatim}
  (defun comp-perm-aux (x y l)
    (if (consp l)
        (cons (nth (nth (car l) y) x)
              (comp-perm-aux x y (cdr l)))
      ()))
  (defund comp-perm (x y n)
    (comp-perm-aux x y (ninit n)))
\end{verbatim}
\end{small}
The behavior of a product of permutations {\tt x} and {\tt y} is characterized by the following:
\begin{small}
\begin{verbatim}
(defthm nth-comp-perm
  (implies (and (posp n) (natp k) (< k n))
           (equal (nth k (comp-perm x y n)) (nth (nth k y) x))))
\end{verbatim}
\end{small}
More generally, we define the product of a list of permutations:
\begin{small}
\begin{verbatim}
  (defun comp-perm-list (l n)
    (if (consp l)
        (comp-perm (car l) (comp-perm-list (cdr l) n) n)
      (ninit n)))
\end{verbatim}
\end{small}
The inverse operator is defined using the function {\tt index} \cite[Sec. 1]{part2}, which gives the location of a member of a list:
\begin{small}
\begin{verbatim}
  (defun inv-perm-aux (x l)
    (if (consp l)
        (cons (index (car l) x) (inv-perm-aux x (cdr l)))
      ()))
  (defund inv-perm (x n)
    (inv-perm-aux x (ninit n)))
\end{verbatim}
\end{small}
It is easily shown that {\tt (slist n)} is a dlist and that {\tt (car (slist n)) = (ninit n)} is a left identity.  After establishing closure, associativity, and the inverse property, we invoke the {\tt defgroup} macro to construct the group:
\begin{small}
\begin{verbatim}
  (defgroup sym (n)
    (posp n)           ;parameter constraint
    (slist n)          ;element list
    (comp-perm x y n)  ;group operation
    (inv-perm x n))    'inverse operator
\end{verbatim}
\end{small}
The length of {\tt (slist n)} is easily computed:
\begin{small}
\begin{verbatim}
  (defthmd order-sym (implies (posp n) (equal (order (sym n)) (fact n))))
\end{verbatim}
\end{small}

\subsection{Transpositions}

A {\it transposition} is a permutation in {\tt (sym n)} that interchanges two indices and leaves all others fixed.  The function {\tt transpose} constructs the transposition of given indices {\tt i} and {\tt j}:
\begin{small}
\begin{verbatim}
  (defun transpose-aux (i j l)
    (if (consp l)
        (if (equal (car l) i)
            (cons j (transpose-aux i j (cdr l)))
          (if (equal (car l) j)
              (cons i (transpose-aux i j (cdr l)))
            (cons (car l) (transpose-aux i j (cdr l)))))
      ()))
  (defund transpose (i j n) (transpose-aux i j (ninit n)))
\end{verbatim}
\end{small}
We define a predicate that characterizes suitable arguments of {\tt transpose}:
\begin{small}
\begin{verbatim}
  (defun trans-args-p (i j n)
    (and (posp n) (natp i) (natp j) (< i n) (< j n) (not (= i j))))
\end{verbatim}
\end{small}
A transposition is a group element of ord 2:
\begin{small}
\begin{verbatim}
  (defthmd transpose-involution
    (implies (trans-args-p i j n)
             (equal (comp-perm (transpose i j n) (transpose i j n) n)
                    (ninit n))))
\end{verbatim}
\end{small}
We need a predicate that recognizes a permutation as a transposition when the interchanged indices are unknown. First we define a function that identifies the least index that is not fixed by a given non-trivial permutation {\tt p}:
\begin{small}
\begin{verbatim}
  (defun least-moved-aux (p k)
    (if (and (consp p) (equal (car p) k))
        (least-moved-aux (cdr p) (1+ k))
      k))
  (defund least-moved (p) (least-moved-aux p 0))
\end{verbatim}
\end{small}
The following predicate recognizes a transposition {\tt p} in {\tt (sym n)}:
\begin{small}
\begin{verbatim}
  (defund transp (p n)
    (let ((m (least-moved p)))
      (and (trans-args-p m (nth m p) n)
           (equal p (transpose m (nth m p) n)))))
  (defthmd transp-transpose
    (implies (trans-args-p i j n)
             (transp (transpose i j n) n)))
\end{verbatim}
\end{small}
A list of transpositions:
\begin{small}
\begin{verbatim}
  (defun trans-list-p (l n)
    (if (consp l)
        (and (transp (car l) n) (trans-list-p (cdr l) n))
      t))
\end{verbatim}
\end{small}
We shall show that every element {\tt p} of {\tt (sym n)} is the product of a list {\tt (trans-list p n)} of transpositions.  Let {\tt m = (least-moved p)}, {\tt q = (transpose m (nth m p) n)}, and {\tt p\$ = (comp-perm q p n)}.  Note that {\tt (least-moved q) = m}.  Therefore, if {\tt k} $<$ {\tt m}, then by {\tt nth-comp-perm},
\begin{small}
\begin{verbatim}
    (nth k p$) = (nth (nth k p) q) = (nth k q) = k,
\end{verbatim}
\end{small}
while
\begin{small}
\begin{verbatim}
    (nth m p$) = (nth (nth m p) q) = m.
\end{verbatim}
\end{small}
Thus, {\tt (least-moved p\$)} $>$ {\tt m}.  This provides a measure for the following recursive definition:
\begin{small}
\begin{verbatim}
  (defun trans-list (p n)
    (declare (xargs :measure (nfix (- n (least-moved p)))))
    (let* ((m (least-moved p))
           (q (transpose m (nth m p) n))
             (p$ (comp-perm trans p n)))
      (if (and (posp n)
               (in p (sym n))
               (< m n))
          (cons trans (trans-list comp n))
        ())))
\end{verbatim}
\end{small}
The desired result is easily proved using the induction scheme provided by the above definition:
\begin{small}
\begin{verbatim}
  (defthmd perm-prod-trans
    (implies (and (posp n) (in p (sym n)))
             (and (trans-list-p (trans-list p n) n)
                  (equal (comp-perm-list (trans-list p n) n)
                         p))))
\end{verbatim}
\end{small}

\subsection{Parity}

An element {\tt p} of {\tt (sym n)} may be represented in various ways as lists of transpositions, but we shall show that {\tt p} determines whether the length of such a list is even or odd.

An {\it inversion} of {\tt p} is a pair of indices {\tt (i . j)} such that {\tt 0} $\leq$ {\tt i} $<$ {\tt j} $<$ {\tt n} and {\tt (nth i p)} $>$ {\tt (nth j p)}.  The {\it parity} of {\tt p} is defined to be that of the number of its inversions.  The formal definition is based on the list {\tt (pairs n)} of pairs {\tt (i . j)} such that {\tt 0} $\leq$ {\tt i} $<$ {\tt j} $<$ {\tt n}.  We extract from this list the list of inversions of {\tt p}:
\begin{small}
\begin{verbatim}
  (defun invs-aux (p pairs)
    (if (consp pairs)
        (if (> (nth (caar pairs) p) (nth (cdar pairs) p))
            (cons (car pairs) (invs-aux p (cdr pairs)))
          (invs-aux p (cdr pairs)))
      ()))
  (defund invs (p n) (invs-aux p (pairs n)))
\end{verbatim}
\end{small}
We now define the parity of {\tt p}:
\begin{small}
\begin{verbatim}
  (defund parity (p n) (mod (len (invs p n)) 2))
\end{verbatim}
\end{small}
Accordingly, {\tt p} is either {\it even} or {\it odd}:
\begin{small}
\begin{verbatim}
  (defund even-perm-p (p n) (equal (parity p n) 0))
  (defund odd-perm-p (p n) (equal (parity p n) 1))
\end{verbatim}
\end{small}
If {\tt p} inverts {\tt i} and {\tt j}, then its inverse {\tt (inv-perm p n)} inverts {\tt (nth j p)} and {\tt (nth i p)}.  It follows that {\tt p} and {\tt (inv-perm p n)} have the same number of inversions and therefore the same parity:
\begin{small}
\begin{verbatim}
  (defthmd parity-inv
    (implies (and (posp n) (in p (sym n)))
             (equal (parity (inv-perm p n) n)
                    (parity p n))))
\end{verbatim}
\end{small}

The proof of the following formula for the parity of a product of permutations is more difficult and is omitted here (see the exposition in the comments in the proof script {\tt groups/symmetric.lisp}):
\begin{small}
\begin{verbatim}
  (defthmd parity-comp-perm
    (implies (and (posp n) (in p (sym n)) (in r (sym n)))
             (equal (parity (comp-perm r p n) n)
                    (mod (+ (parity p n) (parity r n)) 2))))
\end{verbatim}
\end{small}
It follows from {\tt parity-inv} and {\tt parity-comp-perm} that parity is preserved by conjugation:
\begin{small}
\begin{verbatim}
(defthmd parity-conjugate
  (implies (and (posp n)
                (in p (sym n))
                (in a (sym n)))
           (equal (parity (conj p a (sym n)) n)
                  (parity p n))))
\end{verbatim}
\end{small}
Note that a transposition of adjacent indices, {\tt (transpose i (1+ i) n)}, has exactly one inversion and is therefore odd, and every transposition is a conjugate of such a transposition:
\begin{small}
\begin{verbatim}
  (defthmd transpose-conjugate
    (implies (and (trans-args-p i j n) (< (1+ i) j))
             (equal (transpose i j n)
                    (comp-perm (transpose (1+ i) j n)
                               (comp-perm (transpose i (1+ i) n)
                                          (transpose (1+ i) j n)
                                          n)
                               n))))
\end{verbatim}
\end{small}
We may conclude that every transposition is odd:
\begin{small}
\begin{verbatim}
  (defthmd transp-odd (implies (transp p n) (odd-perm-p p n)))
\end{verbatim}
\end{small}
It follows that the parity of a product of a list of transpositions is that of the length of the list:
\begin{small}
\begin{verbatim}
  (defthmd parity-trans-list
    (implies (and (posp n) (trans-list-p l n))
             (equal (parity (comp-perm-list l n) n) (mod (len l) 2))))
\end{verbatim}
\end{small}
In particular, this holds for the canonical factorization:
\begin{small}
\begin{verbatim}
  (defthmd parity-len-trans-list
    (implies (and (posp n) (in p (sym n)))
             (equal (parity p n) (mod (len (trans-list p n)) 2))))
\end{verbatim}
\end{small}

\subsection{Alternating Groups}

The alternating group {\tt (alt n)} is the subgroup of the symmetric group comprising the even permutations:
\begin{small}
\begin{verbatim}
  (defun even-perms-aux (l n)
    (if (consp l)
        (if (even-perm-p (car l) n)
            (cons (car l) (even-perms-aux (cdr l) n))
          (even-perms-aux (cdr l) n))
      ()))
  (defund even-perms (n) (even-perms-aux (slist n) n))
  (defsubgroup alt (n) (sym n)
    (posp n)
    (even-perms n))
\end{verbatim}
\end{small}
It follows from {\tt parity-conjugate} that {\tt (alt n)} is a normal subgroup of {\tt (sym n)}:
\begin{small}
\begin{verbatim}
  (defthmd alt-normal
    (implies (posp n) (normalp (alt n) (sym n))))
\end{verbatim}
\end{small}
Assume {\tt n} $>$ 1 and let {\tt s = (transpose 0 1 n)}, Then {\tt  s} is an odd permutation, and for any odd {\tt p}, {\tt (comp-perm s p n)} is even and therefore {\tt p} belongs to {\tt (lcoset s (alt n) (sym n))}.  Thus, {\tt (alt n)} has only two left cosets:
\begin{small}
\begin{verbatim}
  (defthmd subgroup-index-alt
    (implies (and (natp n) (> n 1))
             (equal (subgroup-index (alt n) (sym n)) 2)))
  (defthmd order-alt
    (implies (and (natp n) (> n 1))
             (equal (order (alt n)) (/ (fact n) 2))))
\end{verbatim}
\end{small}

\section{Group Actions}\label{actions}

\subsection{Definition and the {\tt defaction} Macro}\label{defaction}

Informally, an {\it action} of a group {\tt g} on a dlist {\tt d} is a mapping {\tt a} that assigns to each element {\tt x} of {\tt g} and each member {\tt s} of {\tt d} a member {\tt (act x s a g)} of {\tt d}, satisfying two properties:
\begin{small}
\begin{verbatim}
       (act (e g) s a g) = s
       (act x (act y s a g) a g) = (act (op x y g) s a g).
\end{verbatim}
\end{small}
This may be viewed as a generalization of the operation of a group, which is an action of the group on its own element list.  In fact, we use the same scheme for representing a group action as in the definition of a group: we define an action to be a matrix {\tt a} of members of {\tt d}, the first row of which is {\tt d}, the {\it domain} of {\tt a}:
\begin{small}
\begin{verbatim}
  (defmacro dom (a) `(car ,a))
\end{verbatim}
\end{small}
The ith row of {\tt a} defines the action of the ith element of {\tt g} on the members of {\tt d}:
\begin{small}
\begin{verbatim}
  (defund act (x s a g) (nth (ind s a) (nth (ind x g) a)))
\end{verbatim}
\end{small}
where, according to the definition of {\tt ind} \cite[Sec. 1]{part2},  {\tt (ind s a) = (index s (dom a))}.  Note that the first property above is automatic:
\begin{small}
\begin{verbatim}
  (act (e g) s a g) = (nth (ind s a) (nth 0 a)) = (nth (index s (dom a)) (dom a)) = a.
\end{verbatim}
\end{small}
The defining predicate for an action calls the predicates {\tt aclosedp} and {\tt aassocp}, which exhaustively check the closure and associativity properties:
\begin{small}
\begin{verbatim}
  (defund actionp (a g)
    (and (groupp g)
         (dlistp (dom a))
         (consp (dom a))
         (matrixp a (order g) (len (dom a)))
         (aclosedp a g)
         (aassocp a g)))
\end{verbatim}
\end{small}
It is easily verified that every group is indeed also an action:
\begin{small}
\begin{verbatim}
  (defthm actionp-groupp (implies (groupp g) (actionp g g)))
\end{verbatim}
\end{small}
We have defined a macro for defining parametrized group actions, similar to {\tt defgroup}:
\begin{center}
{\tt (defaction} {\it name args grp cond elts act}{\tt )},
\end{center}
where
\begin{itemize}
\item {\it grp} is the acting group;
\item {\it cond} is a constraint that must be satisfied by the arguments {\it args};
\item {\it elts} is the domain;
\item {\it act} is a term that specifies the action of a group element {\tt x} on a member {\it s} of {\it elts}.
\end{itemize}
Conjugation is an important example.  The form
\begin{small}
\begin{verbatim}
  (defaction conjugacy () g t (elts g) (conj s x g))
\end{verbatim}
\end{small}
constructs the action {\tt conjugacy} and proves three lemmas:
\begin{small}
\begin{verbatim}
  (DEFTHM ACTIONP-CONJUGACY
    (IMPLIES (GROUPP G) (ACTIONP (CONJUGACY G) G)))
  (DEFTHM CONJUGACY-DOM
    (IMPLIES (GROUPP G)
             (EQUAL (DOM (CONJUGACY G)) (ELTS G))))
  (DEFTHM CONJUGACY-ACT-REWRITE
    (IMPLIES (AND (GROUPP G) (IN X G) (IN S (CONJUGACY G)))
             (EQUAL (ACT X S (CONJUGACY G) G) (CONJ S X G))))
\end{verbatim}
\end{small}

An action of a group {\tt g} induces an action by any subgroup of {\tt g}:
\begin{small}
\begin{verbatim}
  (defaction subaction (a g) h
    (and (actionp a g) (subgroupp h g))
    (dom a)
    (act x s a g))
\end{verbatim}
\end{small}
As another example, if h is a subgroup of g, then we have an action of {\tt g} on the left cosets of {\tt h}, characterized by
\begin{small}
\begin{verbatim}
  (act x s (act-lcosets h g) g) = (lcoset (op x (car s) g) h g),
\end{verbatim}
\end{small}
which is again constructed by {\tt defaction}:
\begin{small}
\begin{verbatim}                       
  (defaction act-lcosets (h) g
    (subgroupp h g)
    (lcosets h g)
    (lcoset (op x (car s) g) h g))
\end{verbatim}
\end{small}

\subsection{Orbits and Stabilizers}

If {\tt s} is in the domain of an action {\tt a}, the {\it orbit} of {\tt s} is the ordered list of all {\tt r} in {\tt (dom a)} such that {\tt r = (act x s a g)} for some {\tt x} in {\tt g}:
\begin{small}
\begin{verbatim}
  (defun orbit-aux (s a g l)
    (if (consp l)
        (let ((val (act (car l) s a g))
              (res (orbit-aux s a g (cdr l))))
          (if (member-equal val res)
              res
            (insert val res a)))
      ()))
  (defund orbit (s a g) (orbit-aux s a g (elts g)))
\end{verbatim}
\end{small}
We also define the list of all orbits of an action:
\begin{small}
\begin{verbatim}
  (defun orbits-aux (a g l)
    (if (consp l)
        (let ((res (orbits-aux a g (cdr l))))
          (if (member-list (car l) res)
              res
            (cons (orbit (car l) a g) res)))
      ()))
  (defund orbits (a g) (orbits-aux a g (dom a)))
\end{verbatim}
\end{small}
It is easily shown that every element of the domain belongs to its own orbit and that intersecting orbits are equal (i.e., distinct orbits are disjoint).  It follows that appending the list of orbits yields a permutation of the domain:
\begin{small}
\begin{verbatim}
  (defthmd append-list-orbits
    (implies (actionp a g) (permp (append-list (orbits a g)) (dom a))))
\end{verbatim}
\end{small}
Note that in the case of the {\tt conjugacy} action, the orbit of a group element {\tt x} is the conjugacy class {\tt (conjs x g)} and the class equation \cite[Sec. 8]{part1} is a special case of {\tt append-list-orbits}.

The {\it stabilizer} of an element {\tt s} of the domain of {\tt a} is the ordered subgroup of {\tt g} comprising all {\tt x} such that {\tt (act x s a g) = s}:
\begin{small}
\begin{verbatim}
  (defun stab-elts-aux (s a g l)
    (if (consp l)
        (if (equal (act (car l) s a g) s)
            (cons (car l) (stab-elts-aux s a g (cdr l)))
          (stab-elts-aux s a g (cdr l)))
      ()))
  (defund stab-elts (s a g) (stab-elts-aux s a g (elts g)))
  (defsubgroup stabilizer (s a) g
    (and (actionp a g) (member-equal s (dom a)))  ;constraints
    (stab-elts s a g))                            ;domain
\end{verbatim}
\end{small}

If {\tt r} is in the orbit of {\tt s}, then for some {\tt x} in {\tt g}, {\tt (act x s a g) = r}.  The function {\tt actor} is defined to produce such a value {\tt x}.  This gives rise to the following functions, which may be shown to be inverse bijections between {\tt(lcosets (stabilizer s a g) g)} and {\tt (orbit s a g)}:
\begin{small}
\begin{verbatim}
  (defund lcosets2orbit (c s a g) (act (car c) s a g))
  (defund orbit2lcosets (r s a g) (lcoset (actor r s a g) (stabilizer s a g) g))
\end{verbatim}
\end{small}
Therefore, the lengths of these two lists are equal, and the following is a consequence of {\tt lagrange}:
\begin{small}
\begin{verbatim}
  (defthmd stabilizer-orbit
    (implies (and (actionp a g) (in s a))
             (equal (* (order (stabilizer s a g)) (len (orbit s a g)))
                    (order g))))
\end{verbatim}
\end{small}
Note that the centralizer of a group element {\tt x} is its stabilizer under the {\tt conjugacy} action, and the lemma {\tt len-conjs-cosets} \cite[Sec. 8]{part1} is a case of {\tt stabilizer-orbit}.

\subsection{Conjugation of Subgroups}

The {\it conjugate} of a subgroup {\tt h} of {\tt g} by an element {\tt a} of {\tt g} is a subgroup comprising all conjugates of elements of {\tt h} by {\tt a}.  We define this subgroup to have an ordered element list with respect to {\tt g}, so that element lists of distinct conjugate subgroups cannot be permutations of one another:
\begin{small}
\begin{verbatim}
  (defun conj-sub-list-aux (l a g)
    (if (consp l)
        (insert (conj (car l) a g)
                (conj-sub-list-aux (cdr l) a g)
                g)
      ()))
  (defund conj-sub-list (h a g) (conj-sub-list-aux (elts h) a g))
  (defsubgroup conj-sub (h a) g
    (and (subgroupp h g) (in a g))
    (conj-sub-list h a g))
\end{verbatim}
\end{small}
It is clear that a conjugate of {\tt h} has the same order as {\tt h}, and therefore if one conjugate is a subgroup of another, then they are equal.  Since {\tt h} itself need not be ordered with respect to  {\tt g}, {\tt h} may not be a conjugate of itself, but the conjugate of {\tt h} by {\tt (e g)} (or by any element of {\tt h}, for that matter) has the same elements as {\tt h}:
\begin{small}
\begin{verbatim}
  (defthmd permp-conj-sub-e
    (implies (subgroupp h g)
             (permp (elts (conj-sub h (e g) g)) (elts h))))
\end{verbatim}
\end{small}

Subgroup conjugation is an important example of a group action, the domain of which is a list of all conjugates of a given subgroup {\tt h} of {\tt g}:
\begin{small}
\begin{verbatim}
  (defun conjs-sub-aux (h g l)
    (if (consp l)
        (let ((c (conj-sub h (car l) g))
              (res (conjs-sub-aux h g (cdr l))))
            (if (member-equal c res)
              res
            (cons c res)))
      ()))
  (defund conjs-sub (h g) (conjs-sub-aux h g (elts g)))
  (defaction conj-sub-act (h) g (subgroupp h g) (conjs-sub h g) (conj-sub s x g))
\end{verbatim}
\end{small}

We define the {\it normalizer} of a subgroup {\tt h} of {\tt g} to be the stabilizer of {\tt (conj-sub h (e g) g)}:
\begin{small}
\begin{verbatim}
  (defund normalizer (h g)
    (stabilizer (conj-sub h (e g) g)
                (conj-sub-act h g)
                g))
\end{verbatim}
\end{small}
A subgroup is a normal subgroup of its normalizer:
\begin{small}
\begin{verbatim}
  (defthmd normalizer-normp
    (implies (subgroupp h g) (normalp h (normalizer h g))))
\end{verbatim}
\end{small}
By {\tt stabilizer-orbit}, the number of conjugates of {\tt h} is the index of its normalizer, and therefore, {\tt h} is normal in {\tt g} iff {\tt (mormalizer h g) = g}:
\begin{small}
\begin{verbatim}
  (defthmd index-normalizer
    (implies (subgroupp h g)
             (equal (len (conjs-sub h g))
                    (subgroup-index (normalizer h g) g))))
\end{verbatim}
\end{small}
The normalizer of a conjugte of {\tt h} is a conjugate of the normalizer of {\tt h}:
\begin{small}
\begin{verbatim}
  (defthmd normalizer-conj-sub
    (implies (and (subgroupp m g)
                  (member-equal c (conjs-sub m g)))
             (equal (normalizer c g)
                    (conj-sub (normalizer m g) (conjer-sub c m g) g))))
\end{verbatim}
\end{small}

\subsection{Induced Homomorphism into the Symmetric Group}\label{actsym}

An action {\tt a} of a group {\tt g} associates each element of {\tt g} with a permutation of {\tt (dom a)}.  By identifying an element of {\tt (dom a)} with its index in the list, we have an element of the symmetric group {\tt (sym n)}, where {\tt n = (len (dom a))}.  If {\tt x} is in {\tt g} and {\tt p} is the element of {\tt (sym n)} corresponding to {\tt x}, then for 0 $\leq$ {\tt k} $<$ n, the image of {\tt k} under {\tt p}, {\tt (nth k p)}, is computed by the following:
\begin{small}
\begin{verbatim}
  (defund act-perm-val (x k a g)
    (index (act x (nth k (dom a)) a g)
           (dom a)))
\end{verbatim}
\end{small}
Thus, the element of {\tt (sym n)} corresponding to {\tt x} may be computed recursively:
\begin{small}
\begin{verbatim}
  (defun act-perm-aux (x k a g)
    (if (zp k)
        ()
      (append (act-perm-aux x (1- k) a g)
              (list (act-perm-val x (1- k) a g)))))
  (defund act-perm (x a g) (act-perm-aux x (order a) a g))
  
  (defthmd act-perm-is-perm
    (implies (and (actionp a g) (in x g))
             (in (act-perm x a g) (sym (len (dom a))))))
  (defthm act-perm-val-is-val
    (implies (and (actionp a g) (in x g) (member-equal k (ninit (order a))))
           (equal (nth k (act-perm x a g))
                  (act-perm-val x k a g))))
\end{verbatim}
\end{small}
It is clear that the identity of {\tt g} corresponds to the identity of {\tt (sym n)}, and that the group operation is preserved by this correspondence.  Thus, we have a homomorphism from {\tt g} into the symmetric group:
\begin{small}
\begin{verbatim}
  (defmap act-sym (a g)
    (elts g)
    (act-perm x a g))
  (defthmd homomorphismp-act-sym
    (implies (actionp a g) (homomorphismp (act-sym a g) g (sym (order a)))))
\end{verbatim}
\end{small}
The kernel of {\tt (act-sym a g)} consists of the elements of {\tt g} that act trivially on every element of {\tt(dom a)}.

We have observed that every group {\tt g} is an action of itself on its element list, with {\tt (act g x s g) = (op x s g)}.  The identity of {\tt g} is the only element that acts trivially on every element (or, indeed, on any element).  Therefore, every group {\tt g} is isomorphic to a subgroup of {\tt (sym (order g))}:
\begin{small}
\begin{verbatim}
  (defthm endomorphismp-act-sym-g
    (implies (groupp g) (endomorphismp (act-sym g g) g (sym (order g)))))
\end{verbatim}
\end{small}

As another example, recall the action {\tt act-lcosets} of a group {\tt g} on the left cosets of a subgroup {\tt h}, (Subsection~\ref{defaction}).  Clearly, the kernel of the homomorphism induced by this action is a subgroup of {\tt h}:
\begin{small}
\begin{verbatim}
  (defthmd subgroup-kernel-act-cosets
    (implies (subgroupp h g)
             (subgroupp (kernel (act-sym (act-lcosets h g) g)
                                (sym (subgroup-index h g))
                                g)
                        h)))
\end{verbatim}
\end{small}
This result has the following important consequence:  If {\tt p} is the least prime dividing the order of {\tt g} and {\tt h} is a subgroup of index {\tt p}, then {\tt h} is normal in {\tt g}:
\begin{small}
\begin{verbatim}
  (defthmd index-least-divisor-normal
    (implies (and (subgroupp h g)
                  (> (order g) 1)
                  (equal (subgroup-index h g)
                         (least-prime-divisor (order g))))
             (normalp h g)))
\end{verbatim}
\end{small}
The proof of this theorem also requires the observation that every homomorphism induces an endomorphism on the quotient of its kernel:
\begin{small}
\begin{verbatim}
  (defmap quotient-map (map g h)
    (lcosets (kernel map h g) g)
    (mapply map (car x)))
  (defthmd endomorphismp-quotient-map
    (implies (homomorphismp map g h)
             (endomorphismp (quotient-map map g h) (quotient g (kernel map h g)) h)))
\end{verbatim}
\end{small}
Tho prove {\tt index-least-divisor-normal}, let
\begin{small}
\begin{verbatim}
    k = (kernel (act-sym (act-cosets h g) g) (sym (subgroup-index h g)) g).
\end{verbatim}
\end{small}
We need only show that {\tt k} and {\tt h} have the same elements.  (Since {\tt h} need not be ordered with respect to {\tt g}, the two subgroups may not be equal, but this will be sufficient to conclude that {\tt h} is normal.)  By {\tt endomorphismp-quotient-map}, {\tt (quotient g k)} is isomorphic to a subgroup of {\tt (sym p)}, and therefore {\tt (subgroup-index k g)} divides {\tt (fact p)}, which implies {\tt (subgroup-index k h)} divides {\tt (fact (1- p))}.  If {\tt (subgroup-index k h)} $>$ 1, then {\tt (subgroup-index k g)} has a prime divisor {\tt q}.  Since {\tt q} divides {\tt (fact (1- p))}, {\tt q} $<$ {\tt p}.  But since {\tt q} divides {\tt (order g)}, {\tt q} $\geq$ {\tt p} by assumption, a contradiction.  Thus, {\tt (subgroup-index k h)} = 1, which implies {\tt (permp (elts k) (elts h))}.

\section{Sylow Theorems}\label{sylow}

The {\t Sylow} theorems are a set of related results that provide information pertaining to the number of subgroups of prime power order of a finite group and the relations among them.  These theorems form an important part of group theory, playing a critical role in the classification of finite groups.

Among these results is the statement that the order of a maximal {\tt p}-subgroup of a finite group {\tt g} is the maximal power of {\tt p} that divides the order of {\tt g}.  As a first step, we shall prove that if {\tt h} is a {\tt p}-subgroup of {\tt g} and {\tt p} divides {\tt (subgroup-index p (normalizer h g))}, then {\tt h} is a proper subgroup of a larger {\tt p}-subgroup of {\tt g}, which may be constructed by first applying {\tt cauchy} to construct a subgroup of {\tt (quotient (normalizer h g) h)} of order {\tt p} and then lifting it to {\tt g}:

\begin{small}
\begin{verbatim}
  (defund extend-p-subgroup (h g p)
    (lift (cyclic (elt-of-ord p (quotient (normalizer h g) h))
                  (quotient (normalizer h g) h))
          h
          (normalizer h g)))
  (defthmd order-extend-p-subgroup
    (implies (and (subgroupp h g)
                  (posp n)
                  (elt-of-ord n (quotient (normalizer h g) h)))
             (let ((k (extend-p-subgroup h g n)))
               (and (subgroupp h k)
                    (subgroupp k g)
                    (equal (order k) (* n (order h)))))))
\end{verbatim}
\end{small}
We recursively define a {\tt p}-subgroup {\tt m} = {\tt (sylow-subgroup g p)} of {\tt g} such that {\tt p} does not divide the index of {\tt m} in its normalizer:
\begin{small}
\begin{verbatim}
  (defun sylow-subgroup-aux (h g p)
    (declare (xargs :measure (nfix (- (order g) (order h)))))
    (if (and (subgroupp h g) (primep p)
             (divides p (subgroup-index h (normalizer h g))))
        (sylow-subgroup-aux (extend-p-subgroup h g p) g p)
      h))
  (defund sylow-subgroup (g p) (sylow-subgroup-aux (trivial-subgroup g) g p))
\end{verbatim}
\end{small}
\begin{small}
\begin{verbatim}
  (defthm index-sylow-subgroup
    (implies (and (groupp g) (primep p))
             (let ((m (sylow-subgroup g p)))
               (and (subgroupp m g)
                    (p-groupp m p)
                    (not (divides p (subgroup-index m (normalizer m g))))))))
\end{verbatim}
\end{small}
We aim to show that {\tt p} does not divide the index of {\tt m} in {\tt g}, i.e., {\tt (order m)} is the maximal power of {\tt p} that divides {\tt (order g)}.  To this end, consider the action of {\tt g} on the list of conjugates of {\tt m}.  This action has one orbit, the order of which is the index of the normalizer of {\tt m}.  We shall show that this index is congruent to 1 modulo {\tt p}, and therefore not divisible by {\tt p}.

Consider the restriction of this action to some {\tt p}-subgroup {\tt h} of {\tt g}.  Let {\tt c} be a conjugate of {\tt m}.  By {\tt normalizer-conj-sub}, the normalizer of {\tt c} is a conjugate of the normalizer of {\tt m}, and therefore the index of {\tt c} in {\tt (normalizer c g)} is not divisible by {\tt p}.

Suppose {\tt x} is an element of both {\tt h} and {\tt (normalizer c g)}, but not an element of {\tt c}.  Since the order of {\tt x} in {\tt g} is a power of {\tt p}, the order of the coset of {\tt x} in {\tt (quotient (normalizer c g) c)} is also a power of {\tt p}, and {\tt p} must divide {\tt (subgroup-index c (normalizer c g))}, a contradiction.  Thus, {\tt x} is in {\tt h}, then {\tt x} is in {\tt (normalizer c g)} iff {\tt x} is in {\tt c}.

By {\tt stabilizer-orbit}, the length of the orbit of c under conjugation by {\tt h} is 1 if {\tt h} stabilizes {\tt c}, and otherwise is divisible by {\tt p}.  But {\tt h} stabilizes {\tt c} iff {\tt h} is a subgroup of {\tt (normalizer c g)}, and according to the above observation, this holds iff {\tt h} is a subgroup of {\tt c}:
\begin{small}
\begin{verbatim}
  (defthmd orbit-subaction-div-p
    (implies (and (subgroupp m g)
                  (primep p)
                  (p-groupp m p)
                  (not (divides p (subgroup-index m (normalizer m g))))
                  (subgroupp h g)
                  (p-groupp h p)
                  (in c (conj-sub-act m g)))
             (if (subgroupp h c)
                 (equal (len (orbit c (subaction (conj-sub-act m g) g h) h)) 1)
               (divides p (len (orbit c (subaction (conj-sub-act m g) g h) h))))))
\end{verbatim}
\end{small}

We first apply the above result to the case {\tt h} = {\tt m}.  Since {\tt m} is a subgroup of exactly 1 conjugate of {\tt m}, there is exactly 1 orbit of length 1 and all others have length divisible by {\tt p}:
\begin{small}
\begin{verbatim}
  (defthmd orbit-subaction-m-len-1
    (implies (and (subgroupp m g)
                  (primep p)
                  (p-groupp m p)
                  (not (divides p (subgroup-index m (normalizer m g))))
                  (in c (conj-sub-act m g)))
             (if (equal c (conj-sub m (e g) g))
                 (equal (len (orbit c (subaction (conj-sub-act m g) g m) m)) 1)
               (divides p (len (orbit c (subaction (conj-sub-act m g) g m) m))))))
\end{verbatim}
\end{small}
Appending all orbits yields the first Sylow theorem:
\begin{small}
\begin{verbatim}
  (defthmd sylow-1
    (implies (and (groupp g) (primep p))
             (let ((m (sylow-subgroup g p)))
               (equal (mod (len (conjs-sub m g)) p)
                      1))))
\end{verbatim}
\end{small}
Since {\tt (len (conjs-sub m g)) = (subgroup-index (normalizer m g) g)}, this length divides {\tt (subgroup-index m g)}:
\begin{small}
\begin{verbatim}
  (defthmd sylow-2
    (implies (and (groupp g) (primep p))
             (let ((m (sylow-subgroup g p)))
               (divides (len (conjs-sub m g))
                        (subgroup-index m g)))))
\end{verbatim}
\end{small}
Since {\tt (len (conjs-sub m g)) = (subgroup-index (normalizer m g) g)} is not divisible by {\tt p}, neither is {\tt (subgroup-index m g)}:
\begin{small}
\begin{verbatim}
  (defthmd sylow-3
    (implies (and (groupp g) (primep p))
             (not (divides p (subgroup-index (sylow-subgroup g p) g)))))
\end{verbatim}
\end{small}

The final Sylow theorem states that every {\tt p}-subgroup of {\tt g} is a subgroup of some conjugate of {\tt m}.  This is another consequence of {\tt orbit-subaction-div-p}: If {\tt h} were a counterexample to this claim, then according to {\tt orbit-subaction-div-p}, the length of every orbit of {\tt h} would be divisible by {\tt p}, contradicting {\tt mod-len-conjs-sub}.

The statement of the theorem requires the following finction, which searches a list {\tt l} of subgroups of {\tt g} for one that contains {\tt h} as a subgroup:
\begin{small}
\begin{verbatim}
  (defun find-supergroup (h l)
    (if (consp l)
        (if (subgroupp h (car l))
            (car l)
          (find-supergroup h (cdr l)))
      ()))
  (defthmd sylow-4
    (implies (and (groupp g) (primep p) (subgroupp h g) (p-groupp h p))
             (let* ((m (sylow-subgroup g p))
                    (k (find-supergroup h (conjs-sub m g))))
               (and (member-equal k (conjs-sub m g))
                    (subgroupp h k)))))
\end{verbatim}
\end{small}

\section{Simple Groups}\label{simple}

A group is {\it simple} if has has no proper normal subgroup.
\begin{small}
\begin{verbatim}
  (defund proper-normalp (h g)
    (and (normalp h g) (> (order h) 1) (< (order h) (order g))))
\end{verbatim}
\end{small}
Simple groups play an important role in the classification of finite groups.  Since every group of prime order is simple, we focus on groups of composite order.  One class of interest is that of the alternating groups.  Note that {\tt (alt 4)} is not simple, as may be verified by direct computation:
\begin{small}
\begin{verbatim}
  (defthmd alt-4-not-simple
    (proper-normalp (subgroup '((0 1 2 3) (1 0 3 2) (2 3 0 1) (3 2 1 0)) (sym 4))
                    (alt 4)))
\end{verbatim}
\end{small}
However, {\tt (alt n)} is simple for all {\tt n} $\geq$ 5.  We shall prove this only for the case {\tt n} = 5: {\tt (alt 5)} is a simple group of order 60.  In contrast to the more general theorem, our proof of this result is largely computational.  We shall also prove, as an illustration of the Sylow theorems, that there are no simple groups of composite order less than 60.

\subsection{Simplicity of {\tt (alt 5)}}

Let {\tt h} be a normal subgroup of {\tt g}.  The function {\tt conjs-list} \cite[Sec. 8]{part1} constructs a list of the non-central conjugacy classes of {\tt g}.  We define {\tt (select-conjs (conjs-list h) h)} to extract the conjugacy classes that are included in {\tt h}:
\begin{small}
\begin{verbatim}
  (defun select-conjs (l h)
    (if (consp l)
        (if (in (caar l) h)
            (cons (car l) (select-conjs (cdr l) h))
          (select-conjs (cdr l) h))
      ()))
\end{verbatim}
\end{small}
if we append the elements of that list together with the elements of {\tt h} that belong to the center of {\tt g}, we have a permutation of {\tt (elts h)}.  In the case of interest the center happens to be trivial.  This gives us an expression for the order of {\tt h}:
\begin{small}
\begin{verbatim}
  (defthmd len-select-conjs
    (implies (and (normalp h g) (equal (cent-elts g) (list (e g))))
             (equal (order h)
                    (1+ (len (append-list (select-conjs (conjs-list g) h)))))))
\end{verbatim}
\end{small}
Thus, {\tt (order h) = (1+ (len (append-list l)))} for some sublist {\tt l} of {\tt (conjs-list g)}.  We need only compute this value for all such sublists of {\tt (conjs-list (alt 5))} and observe that none of these values is a proper divisor of 60.

However, the function {\tt conjs-list} is computationally impractical for a group of order 60.  We define a more efficient and provably equivalent function, {\tt conjs-list-fast}, based on a tail-recursive version of {conjs}.  The lengths of the conjugacy classes of {\tt (alt 5)} can be easily computed using this function:
\begin{small}
\begin{verbatim}
  (defun lens (l)
    (if (consp l)
        (cons (len (car l)) (lens (cdr l)))
      ()))
  (defthmd lens-conjs-list-alt-5
    (equal (lens (conjs-list-fast (alt 5))) '(20 12 12 15)))
\end{verbatim}
\end{small}
Clearly, no list of distinct members of this list has a sum that is a proper divisor of 60.  Once we establish this simple fact (which requires some work), our theorem follows from {\tt lagrange}:
\begin{small}
\begin{verbatim}
  (defthmd alt-5-simple (not (proper-normalp h (alt 5))))
\end{verbatim}
\end{small}

\subsection{Groups of Lesser Order}

For every group {\tt g} of composite {\tt n} $<$ 60, we shall construct a proper normal subgroup of {\tt g}.  We begin with the case of a prime power: {\tt n = (expt p k)}, where {\tt k} $>$ 1.  By {\tt center-p-group} (book {\tt cauchy}), {\tt (order (center g))} $>$ 1.  If {\tt (order (center g))} $<$ {\tt (order g)}, then {\tt (center (g))} is a proper normal subgroup.  In the remaining case, {\tt (center g) = g}, and hence {\tt g} is abelian.  Thus we need only show that {\tt g} has a proper subgroup.  But this follows from {\tt cauchy}, which guarantees an element of order {\tt p}.  This leads to the following definition and lemma:
\begin{small}
\begin{verbatim}
  (defund normal-subgroup-prime-power (p k g)
    (declare (ignore k))
    (if (< (order (center g)) (order g))
        (center g)
      (cyclic (elt-of-ord p g) g)))
  (defthm proper-normalp-prime-power
    (implies (and (groupp g)
                  (equal (order g) (expt p k))
                  (primep p)
                  (natp k)
                  (> k 1))
             (proper-normalp (normal-subgroup-prime-power p k g) g)))
\end{verbatim}
\end{small}

The rest of the proof is based mainly on the Sylow theorems.  We consider various cases according to the prime factorization of {\tt n}.  As a notational convenience, we shall denote {\tt (sylow-subgroup g p)} by {\tt hp} and {\tt (len (conjs-sub hp g))} by {\tt np}.

Suppose {\tt n = (* p q)}, where {\tt p} and {\tt q} are primes and {\tt p} $<$ {\tt q}.  By the Sylow theorems, {\tt nq} divides {\tt p} and {\tt (mod nq p) = 1}.  It follows that {\tt np = 1}, which implies {\tt (sylow-subgroup g q)} is normal in {\tt g}.
\begin{small}
\begin{verbatim}
  (defund normal-subgroup-pq (p q g)
    (declare (ignore p))
    (sylow-subgroup g q))
  (defthm proper-normalp-pq
    (implies (and (groupp g) (equal (order g) (* p q))
                  (primep p) (primep q) (< p q))
             (proper-normalp (normal-subgroup-pq p q g) g)))
\end{verbatim}
\end{small}

Next, we consider the case {\tt n = (* p p q)}, where {\tt p} and {\tt q} are primes.  We must show that either {\tt np = 1} or {\tt nq = 1}.  Suppose not.  Since {\tt np} divides {\tt q} and {\tt (mod np p) = 1}, {\tt q} $>$ {\tt p}.  Since {\tt nq} divides {\tt (* p p)} and {\tt (mod nq q)} = 1, {\tt nq = (* p p)} and {\tt q} divides {\tt (1- (* p p))}.  Thus, {\tt q} divides either {\tt (1- p)} or {\tt (1+ p)}.  Since {\tt q} $>$ {\tt p}, {\tt q = (1+ p)}, which implies {\tt p} = 2, {\tt p} = 3, and {\tt n} = 12.  Since {\tt n3} = 4 and each 3-Sylow subgroup has 2 non-trivial elements, {\tt g} has 8 elements of order 3.  Since {\tt n2} $>$1, {\tt g} has more than 4 elements of order dividing 4, a contradiction.
\begin{small}
\begin{verbatim}
  (defund normal-subgroup-ppq (p q g)
    (if (normalp (sylow-subgroup g p) g)
        (sylow-subgroup g p)
      (sylow-subgroup g q)))
  (defthm proper-normalp-ppq
    (implies (and (groupp g)
                  (equal (order g) (* p p q))
                  (primep p)
                  (primep q)
                  (not (equal p q)))
             (proper-normalp (normal-subgroup-ppq p q g) g)))
\end{verbatim}
\end{small}

There are eight remaining cases, which are treated individually: 24, 30, 36, 40, 42, 48, 54, and 56.  Consider the case {\tt n} = 24.  Assume {\tt n2} $>$1 and let h21 and h22 be distinct members of (conj-subs h2 g).  Then {\tt (order h21) = order h22) = 8}.  Let {\tt k = (group-intersection h21 h22 g)}.  Then {\tt (order k)} $\leq$ 4 and by {\tt len-products} \cite[Sec. 3]{part2},
\begin{center}
   {\tt (len (products h21 h22 g)) = (/ (* (order h1) (order h2)) (order k))} $\leq$ 24,
\end{center}
which implies {\tt (order k)} = 4 and {\tt (len (products h21 h22 g))} = 16.  By {\tt index-least-divisor-normal} (Subsection~\ref{actsym}), {\tt k} is normal in both {\tt h21} and {\tt h22}.  It follows that {\tt (normalizer k g)} contains {\tt (products h21 h22 g)}.  Consequently, {\tt (order (normalizer k g))} $\geq$ 16, which implies {\tt (normalizer k g) = g} and {\tt k} is normal in {\tt g}.  Thus, we have the following:
\begin{small}
\begin{verbatim}
  (defund normal-subgroup-24 (g)
    (let* ((h2 (sylow-subgroup g 2))
           (h21 (car (conjs-sub h2 g)))
           (h22 (cadr (conjs-sub h2 g)))
           (k (group-intersection h21 h22 g)))
      (if (normalp h2 g)
          h2
        k)))
  (defthm proper-normalp-24
    (implies (and (groupp g) (equal (order g) 24))
             (proper-normalp (normal-subgroup-24 g) g)))
\end{verbatim}
\end{small}
We omit the other seven cases, which use the same techniques as illustrated as above.  We combine these results in a function that splits into cases corresponding to the composite integers less than 60:
\begin{small}
\begin{verbatim}
  (defund normal-subgroup (g)
    (case (order g)
      (4 (normal-subgroup-prime-power 2 2 g))
      (6 (normal-subgroup-pq 2 3 g))
      (8 (normal-subgroup-prime-power 2 3 g))
      (9 (normal-subgroup-prime-power 3 2 g))
      (10 (normal-subgroup-pq 2 5 g))
      (12 (normal-subgroup-ppq 2 3 g))
      ...
      (56 (normal-subgroup-56 g))
      (57 (normal-subgroup-pq 3 19 g))
      (58 (normal-subgroup-pq 2 29 g))))

  (defthm no-simple-group-of-composite-order<60
    (implies (and (natp n) (> n 1) (< n 60) (not (primep n))
                  (groupp g) (equal (order g) n))
             (proper-normalp (normal-subgroup g) g)))
\end{verbatim}
\end{small}

\section{Conclusion}

Our survey of this rich topic is far from complete.  We anticipate enhancements of the theory, such as the representation of a permutation as a product of disjoint cycles, which is required, for example, for a general proof of the simplicity of {\tt (alt n)}.  However, the intended scope of the project has essentially been realized.  The combined content of the {\tt groups} directory is a close approximation to that of an advanced undergraduate course that the author taught at The Cooper Union in the Spring of 1976.

The concluding section of Part I discusses the long-term objective of a formalization of algebraic number theory.  The next steps in this direction are elementary linear algebra and Galois theory, the first of which is underway.  We note one important difference between our approaches to groups, on one hand, and fields and vector spaces on the other.  As we have observed, our interest in finite groups and the importance of proof by induction on the order of a group led us away from the characterization of a group by means of encapsulated constrained functions in favor of an explicit defining predicate.  On the other hand, since we are interested in both infinite and finite fields (and the role of induction is less critical even in the latter case), we are instead pursuing the encapsulation approach in the formalization of fields as well as finite dimensional vector spaces..  A progress report may be expected at the next ACL2 workshop.

\nocite{*}
\bibstyle{eptcs}
\bibliographystyle{eptcs}
\bibliography{groups3}
\end{document}